\documentclass[aps,pra,10pt,notitlepage,twocolumn,showpacs]{revtex4-1}
\usepackage[dvips]{graphicx}
\usepackage{amssymb,amsfonts}
\usepackage{amsmath,color}

\begin{document}

\author{P. Sza\'nkowski, M. Trippenbach and J. Chwede\'nczuk}
\affiliation{
  Faculty of Physics, University of Warsaw, ul. Ho\.{z}a 69, PL--00--681 Warszawa, Poland
}

\title{Parameter estimation in a memory-assisted noisy quantum interferometry}


\begin{abstract}
  We demonstrate that memory in an $N$-qubit system subjected to decoherence, is a potential resource for the slow-down of the entanglement decay. 
We show that this effect can be used to retain the sub shot-noise sensitivity of the parameter estimation in quantum interferometry.
  We calculate quantum Fisher information, which sets the ultimate bound for the precision of the estimation. We also derive the sensitivity of such a noisy interferometer, 
  when the phase is either estimated from the measurements of the population imbalance or from the one-body density.
\end{abstract}
\maketitle

\section{Introduction}
Non-classical correlations proved to be a key concept in many areas of modern science, such as quantum information \cite{horo}, quantum computation, cryptography, teleportation
\cite{weedbrook} and even biology \cite{lambert}. In ideal circumstances, non-classical states can be prepared and utilized at will.
For instance, quantum metrology employs correlated states to reach the sub shot-noise (SSN) precision of parameter estimation \cite{cirac,giovanetti,pezze,cronin}.
However, a coupling with environment inevitably
destroys subtle quantum correlations and drives the system into a classical mixture. This process, called  decoherence,
imposes severe limitations on what can be achieved in the laboratory \cite{decoh,open,dobrz}.
Motivated by recent interferometric experiments with ultra-cold atoms \cite{esteve,riedel,gross,smerzi,app},
we study the impact of decoherence on a collection of $N$ bosonic qubits. In particular, we investigate the influence of the decoherence on the performance
of atomic interferometers and
focus on an imprint of a relative phase $\theta$ between the two modes \cite{chwed}. This process lasts for a time $\tau$ and is accompanied by the coupling of the system to environment.
As our main result, we demonstrate -- both analytically and numerically -- that
memory opens new possibilities to counteract the decoherence during the evolution of the two-mode quantum state under the influence of the environment.

This paper is organized as follows. In Section \ref{sec.form} we introduce the system of $N$ qubits coupled to the environment and present a detailed discussion of a physically justified model of noise.
In Section \ref{sec.sol}, using the formalism of the spherical tensors, we dervie the expression for the dynamics of the density matrix. In Section \ref{sec.evo} we analyze the dynamics of the system in
detail and show that memory can preserve subtle quantum correlations. In Section \ref{sec.qfi} we turn our attention to the application of the memory effect in quantum interferometry. We show that inded
memory can be utilized to maintain high precision of an interferometer. In Section \ref{sec.cfi} we analyze the impact of noise on some particular estimation schemes. We conclude in Section \ref{sec.conc}.

\section{Formulation of the problem}
\label{sec.form}
A general approach for finding the dynamics of the system ($S$) interacting with an environment ($E$) relies upon solving the 
equation  for
 the density matrix of the system alone $\hat\varrho_S$, i.e. $i\partial_t\hat\varrho_S=\mathrm{Tr}_E\{[\hat H_{\rm tot},\hat\varrho_{SE}]\}$.
Here we set $\hbar\equiv1$ and $\hat\varrho_{SE}$ is the total density matrix of the system and the environment, while the reduced density matrix of the system
is obtained by tracing out the environmental degrees of freedom $\hat\varrho_S=\mathrm{Tr}_E\{\hat\varrho_{SE}\}$.
 The Hamiltonian $\hat H_{\rm tot}$
is a sum of three parts $\hat H_S$, $\hat H_E$ and $\hat H_{\rm int}$.

We now specify the Hamiltonian of the system $\hat H_S$, and note that
a collection of $N$ bosonic qubits is equivalent to a single spin-$\frac N2$ particle. It can be described by the
angular momentum operators  
\begin{equation}
  \hat J_k= \frac12\sum_{i=1}^N\hat\sigma_k^{(i)}
\end{equation}
where $\hat\sigma^{(i)}_k$ is the $k$-th Pauli matrix ($k=x,y,z$) of the $i$-th particle.
We assume, that the system alone is undergoing interferometric transformation, which is the imprint of the relative phase $\theta$ between the two modes, 
\begin{equation}\label{H_S}
  \hat H_S=\Omega\hat J_z.
\end{equation} 
Here, $\Omega$ is an external field, which is related to the interferometric phase by $\theta=\Omega\tau$. Basing on recent experimental results \cite{riedel, gross}, we assumed that the two-body interactions, which are essential for the state preparation \cite{cirac,pezze,giulia},
are absent during the phase imprint.

In the next step, one should specify the Hamiltonian of the environment $\hat H_E$, and determine how it interacts with the system via $\hat H_{\rm int}$.
However, this {\it ab initio} method  is almost always impractical.
Construction of a realistic model of $\hat H_E$ and $\hat H_{\rm int}$ is difficult because the necessary information
about relations between numerous degrees of freedom is hardly accessible.
Moreover, tracing out the environment is challenging and usually gives a closed equation for $\hat\varrho_S$ (then called the quantum Master Equation \cite{vankampen,leggett}) only upon further radical approximations.
The only known treatable case is the
 spin-boson model \cite{leggett}, which depicts a linear interaction of a single qubit with a collection
of independent harmonic oscillators.

Below we discuss a phenomenological model of the system--environment interaction, which circumvents the difficulty with determining the detailed description of the environment.
A necessary assumption for this construction is that the back-action of the system on the environment is either absent or can be safely neglected. If this is the case,
the only way the system can be affected by the environment is through an effective external field generated by it. The qubits are indistinguishable bosons, thus they occupy the same configuration space and each particle has to be coupled to the exactly the same field ${\mathbf\Omega}_E(t)$. Therefore, the interaction Hamiltonian is 
\begin{equation}
  \hat H_{\rm int}=\frac12{\mathbf\Omega}_E(t)\cdot\sum_{i=1}^N \hat{\boldsymbol\sigma}^{(i)}\equiv{\mathbf\Omega}_E(t)\cdot\hat{\mathbf J}
\end{equation} 
where  $\hat{\boldsymbol\sigma}^{(i)}$ is a three-component vector of Pauli matrices for the $i$-th qubit.

To determine the exact form of the field ${\mathbf\Omega}_E(t)$ is as difficult as finding $\hat H_E$. Nevertheless, statistical
properties of the environment should be much easier to either obtain or deduce. This observation allows for the final step, where we replace ${\mathbf\Omega}_E(t)$ with a fluctuating
field described by a stochastic process ${\boldsymbol{\omega}}(t)$ chosen to reflect the aforementioned properties \cite{gardiner}. 

In this way, the explicit presence of the environmental degrees of freedom is mimicked by the fluctuations of the field and the Hamiltonian $\hat H_{\rm tot}$ is replaced by
\begin{equation}\label{ham}
  \hat H=\Omega\hat J_z+\boldsymbol{\omega}(t)\cdot\hat{\mathbf J}.
\end{equation}
We underline the fact that all qubits evolve under the common noise, because the system is composed of identical bosons. Consequently, a collection of $N$ bosonic qubits is equivalent to a single spin-$\frac N2$ particle due to symmetrization over all permutations of particles, which is enforced by the indistinguishability and is conserved by the dynamics.

The evolution of the system generated by the stochastic Hamiltonian (\ref{ham}) can be understood as follows. 
For every trajectory (realization) of the stochastic process $\boldsymbol\omega_0(t)$ chosen at random with the probability distribution $\mathcal P(\boldsymbol\omega_0)$ 
the Hamiltonian (\ref{ham}) generates a unitary evolution of the initial state. The single-trajectory output state, denoted by $\hat\varrho(t;\boldsymbol\omega_0)$, is given by the formula
\begin{equation}\label{single}
  \hat\varrho(t;\boldsymbol\omega_0) = \hat U(t;\boldsymbol\omega_0)\,\hat\varrho(0)\,\hat U^\dagger(t;\boldsymbol\omega_0).
\end{equation} 
Here, the unitary evolution operator involves the time ordering and reads
\begin{equation}
  \hat U(t;\boldsymbol\omega_0)=\mathcal T\exp\left[-i\int_0^t dt'\left(\Omega \hat J_z +\boldsymbol\omega_0(t')\cdot\hat{\mathbf J}\right)\right].
\end{equation}
Since we do not posses the information about which trajectory governs a given run of the experiment, the density matrix of the system $\hat\varrho_S(t)$ is a mixture of all possible choices
\begin{equation}\label{rho_mix}
  \hat\varrho_S(t)= \sum_{\boldsymbol\omega_0}\hat\varrho(t;\boldsymbol\omega_0) = \int\mathcal\!\!\mathcal D\boldsymbol\omega_0\,\mathcal P(\boldsymbol\omega_0)\,\hat\varrho(t;\boldsymbol\omega_0).
\end{equation}
Here $\int \mathcal D\boldsymbol\omega_0$ is a functional integral over the space of real vectorial functions. 
On the other hand, statistical mixture of all possible realizations of the fluctuating field is the same as the definition of the average over process $\boldsymbol\omega$ of the stochastic density matrix 
$\hat\varrho(t)$ obtained as a solution to the stochastic equation of motion provided by (\ref{ham}). Hence, the system density matrix is also equal to
\begin{equation}
\hat\varrho_S(t) = \overline{\hat\varrho(t)},
\end{equation}
where $\overline{(\ )}$ denotes the average. We recognize in this step an analog of tracing over the environmental degrees of freedom performed in the approach of the Master equation.
In this way, we have constructed the general framework of the system-environment model. To continue further on it is necessary to specify the statistical properties of
$\boldsymbol{\omega}(t)$, i.e. the probability distribution $\mathcal P(\boldsymbol\omega_0)$. 
Below we argue that the stationary Gaussian process is the most natural choice which accounts for the wide variety of physical situations.

Typically, we can expect that the environment consists of many approximately independent sources of fluctuations, which sum up to $\boldsymbol{\omega}(t)$. If this is the case, than
according to the Central Limit Theorem, $\boldsymbol{\omega}(t)$ tends to a Gaussian process with the growing number of sources. Therefore, the process is fully determined by its average,
$\overline{\boldsymbol{\omega}(t)}$, and the correlation function, $\overline{\omega_i(t)\omega_j(t')}=\kappa_{ij}(t,t')$. Here,  subscripts $i,j$ denote orthogonal components
of the vector field $\boldsymbol{\omega}(t)$. Moreover, the environment is expected to be in a stationary state -- for example
in the thermal equilibrium. In line with the ``no-back-action'' postulate which states that the system does not disturb the environment, there is no distinguished instant of
time. Hence, also the process itself must be stationary, which implies that the average is a constant and the correlation function depends only on time difference. To make the
discussion even more transparent, we choose $\boldsymbol{\omega}(t)$ to be isotropic -- which applies for an environment without any distinguished direction. Later we will revise this assumption. To summarize,
\begin{equation}\label{iso}
  \overline{\boldsymbol{\omega}(t)}=0\ \ \ \ \ \mathrm{and}\ \ \ \ \ \overline{\omega_i(t)\omega_j(t')}=\kappa(|t-t'|)\delta_{ij},
\end{equation}
where we set the average to be zero -- without any loss of generality. The strength of the fluctuations is given by the variance  $\omega_0^2=\kappa(0)=\overline{\omega_i^2(t)}$.

In the majority of relevant situations, the correlation function
$\kappa(|t-t'|)$ rapidly tends to zero when $|t-t'|>\tau_c$, where $\tau_c$ is the correlation time, which sets the time scale for the process \cite{kampen_cumulant}.


The non-zero correlation time connects the events from various instants of the evolution in a nontrivial way, i.e. not only through the initial conditions. Therefore, if the evolution is
governed by a processes with $\tau_c\neq0$, we say that the system has memory. Moreover, $\tau_c$ can be considered as a measure of memory, so the larger $\tau_c$, the more memory is
present in the system.
The proposed notion of memory stands in line with the rigorous definition of non-Markovian process: the solution of stochastic equation of motion ($\hat\varrho(t)$ before averaging in our case) is Markovian, i.e. memory-less, iff the noise driving the system has a vanishing correlation time \cite{vankampen_colored}.

This type of Gaussian, colored noise occurs, for example, in the magnetic traps induced by the current running through the coils. The thermal fluctuations of the current result in the Gaussian fluctuations of the magnetic
field with $\omega_0^2$ proportional to the temperature and the correlation time proportional to the capacitance of the coils (parasitic and otherwise).

Finally, we comment on the applicability of the presented model of the system--environment interactions. in some physical cases it might be necessary to include the action of the system on the environment. However, this extension is not necessary, in the light of our main conclusion which states
that quantum correlations and high efficiency of the interferometer can be preserved if the experiment is carried out on the timescale of $\tau_c$. 
The noise affecting the system is generated by a complex evolution of the environment, which in principle could be altered by 
the system.
This disturbance modifies the properties of the noise felt by the system and effectively appears as a self-interaction mediated by the environment,
hence it is a second-order correction and at the time-scale of the correlation time can be safely neglected \cite{szan_rulez}.
Moreover, the approximation of this noise as a Gaussian process, even if the Central Limit Theorem fails,
is justified by noting that higher order correlations come into play for times longer then $\tau_c$ \cite{kampen_cumulant, cywinski_08}.

\section{Stochastic equations of motion}
\label{sec.sol}
The stochastic density matrix satisfies the von Neumann equation of motion $i\partial_t\hat\varrho(t)=[\hat H,\hat\varrho(t)]$, with the Hamiltonian
given by Eq.~(\ref{ham}). When $\hat\varrho(t)$ is spanned by the appropriately chosen basis operators -- which respect the symmetry of the Hamiltonian -- the equations of motion greatly
simplify. In our case this orthonormal basis is composed by a collection of spherical tensor operators \cite{sakurai},
which are irreducible representations of the rotation group and are defined by a following set of conditions
\begin{eqnarray}
  &&\left[\hat J_z,\hat T^{(j)}_m\right]=m\hat T^{(j)}_m\\
  &&\left[\hat J_{\pm},\hat T^{(j)}_m\right]=\sqrt{j(j+1)-m(m\pm1)}\hat T^{(j)}_{m\pm1}\\
  &&{\rm Tr}\left(\hat T^{(j)\dagger}_m\hat T^{(j')}_{m'}\right)=\delta_{jj'}\delta_{mm'},
\end{eqnarray}
where $\hat J_{\pm}=\hat J_x\pm i\hat J_y$.
The density matrix decomposed in the basis of $\hat T^{(j)}_m$'s reads
\begin{equation}\label{den_mat}
  \hat\varrho(t)=\sum_{j=0}^N\sum_{m=-j}^j\ \big\langle\hat T^{(j)}_m(t)\big\rangle^{\!*}\ \hat T^{(j)}_m.
\end{equation}
Only the terms $\big\langle\hat T^{(j)}_m(t)\big\rangle={\rm Tr}\left[\hat T^{(j)}_m(t)\hat\varrho(0)\right]$ are affected by fluctuations,
thus the evolution is fully determined by these expectation values. 
The above parametrization of the density matrix (\ref{den_mat}) might seem complicated, but it gives a particularly simple set of equations
\begin{equation}\label{eq_basis}
  i\partial_t\big\langle\hat T^{(j)}_m(t)\big\rangle=\sum_{m'=-j}^j\!\!\mathbb{H}^{(j)}_{mm'}(t)\big\langle\hat T^{(j)}_{m'}(t)\big\rangle,
\end{equation}
where $\mathbb{H}^{(j)}_{mm'}(t)=\langle j,m'|\hat H|j,m\rangle$ and $|j,m\rangle$ are the eigen-states of the angular momentum operators, which will be consequently used as the basis
states. The formal solution of Eq.~(\ref{eq_basis}) is
given in terms of the time-ordered exponential which acts on the initial expectation values $\big\langle\hat T^{(j)}_{m}\big\rangle=\mathrm{Tr}\left[{\hat T^{(j)}_{m}\hat\varrho(0)}\right]$,
i.e.
\begin{equation}\label{stochastic_tens}
  \big\langle\hat T^{(j)}_m(t)\big\rangle=\sum_{m'}\!\!
  \left[\mathcal{T}\!\exp\!\left(\!-i\!\!\int_0^t\!\!\!dt'\,\mathbb{H}^{(j)}(t')\right)\right]_{mm'}\!\!\!\big\langle\hat T^{(j)}_{m'}\big\rangle.
\end{equation}
The final and the most challenging step is to calculate the average of (\ref{stochastic_tens}) using the ``cumulant expansion'' method described in detail in \cite{kubo}. 
However, even this procedure does not provide an analytical outcome apart from two cases when either the Hamiltonians commute at different times (i.e only fluctuations along $z$-axis are present) 
or when the correlation time vanishes (the Markovian limit).
Still, for Gaussian process an excellent approximation is obtained by restricting the cumulant series to the second order \cite{nota_cut}, and this yields
\begin{equation}\label{tens_sol}
  \big\langle\overline{\hat T^{(j)}_m(t)}\big\rangle=e^{-m^2\Gamma_0-(j(j+1)-m^2)\Gamma_+}e^{-im(\Omega t-\Gamma_-)}\big\langle\hat T^{(j)}_m\big\rangle.
\end{equation}
The decay rates are defined by the function
\begin{equation}
  \Gamma(t,\Omega)=\int_0^t\!dt_1\int_0^{t_1}\!\!dt_2\,\kappa(t_1-t_2)\,e^{i\Omega (t_1-t_2)}
\end{equation}
and read
\begin{eqnarray}
  &&\Gamma_0=\Gamma(t,0)\\
  &&\Gamma_+={\rm Re}[\Gamma(t,\Omega)]\\
  &&\Gamma_-={\rm Im}[\Gamma(t,\Omega)].
\end{eqnarray}
The two rates $\Gamma_{\pm}$ result from the fluctuations in the $x-y$ plane,
while $\Gamma_0$ originates only from the $z$-component of the noise.

From Eq.~(\ref{tens_sol}) it is evident that the density matrix elements are damped by the fluctuations. As we argue below, the rate of damping can be substantially
decreased in the presence of memory.

\section{Evolution of the density matrix}
\label{sec.evo}
We illustrate the memory effect by taking the system to be initially in a pure state $\hat\varrho_S(0)=\big|\Psi\big\rangle\big\langle\Psi\big|$, and choose
$\big|\Psi\big\rangle=\frac1{\sqrt 2}\left(\big|\frac N2,\frac N2\big\rangle+\big|\frac N2,-\frac N2\big\rangle\right)$ to be the maximally entangled ``NOON'' state. When $\hat\varrho_S(0)$
is decomposed into the basis of spherical tensors, it reads
\begin{equation}
  \hat\varrho_S(0)=\sum_{j=0}^{\frac N2}\big\langle\hat T^{(2j)}_0\big\rangle\hat T^{(2j)}_0
  +\frac12\left(\hat T^{(N)}_N+\hat T^{(N)}_{-N}\right).
\end{equation}
The first term is the diagonal of the density matrix spanned on tensors with $m=0$. Away from the diagonal, the only non-zero elements are those in the top (spanned on the $m=N$ tensors)
and the bottom ($m=-N$)
corners. The presence of tensors with $m\neq0$ in the density matrix of the NOON state is crucial for the SSN sensitivity of the $\hat J_z$ interferometer \cite{chwed}.
However, due to damping in Eq.~(\ref{tens_sol}), asymptotically only the scalar term $\hat T^{(0)}_0\propto \hat{\mathbf 1}$ remains nonzero and the density matrix tends to a completely mixed state
$\frac1{N+1}\hat{\mathbf 1}$. It does not depend on $\Omega$ and is useless for parameter estimation.
\begin{figure}[hbt]
  \centering
  \includegraphics[clip,
  scale=0.3]{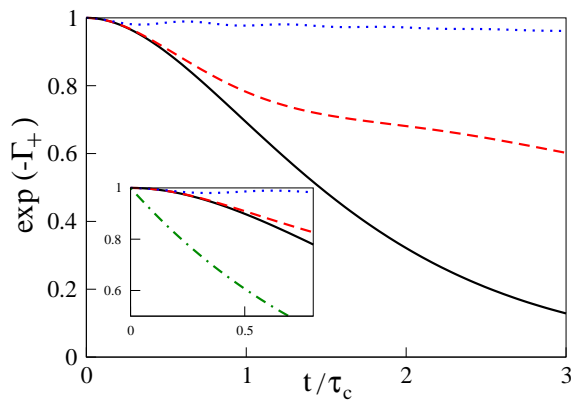}
  \caption{The temporal behavior of the damping factor $\exp\left(-\Gamma_+\right)$ for $\Omega\tau_c=0$ (solid black), $\Omega\tau_c=2.5$ (dashed red)
    and $\Omega\tau_c=10$ (dotted blue).
      The value of $\Gamma_+$ is calculated for the Ornstein-Uhlenbeck process \cite{ornstein}, for which $\kappa(t)=\omega_0^2\exp\left(-\frac{|t|}{\tau_c}\right)$ with $\omega_0\tau_c=1$.
      The inset focuses on early times and compares the colored-noise cases with the white-noise limit $\exp\left(-\omega_0^2\tau_c\,t\right)$ (dot-dashed green).}
  \label{scales}
\end{figure}

In order to appreciate the role of memory as a resource for the decoherence slow-down, we compare the finite correlation time ($\tau_c\neq0$) case with the ``white noise'' limit. This limit is achieved by letting $\tau_c\to0$ while keeping  $\frac 1T\equiv\int_0^\infty \!\kappa(s)ds$ constant. As a result, the correlation function tends to the Dirac delta,
$\kappa(t)\to\frac2T\delta(t)$, which gives $\Gamma(t,\Omega)\to\frac tT$.
It can be related to the colored noise decay rate, which for short times is $\Gamma(t,\Omega)\simeq\frac12\omega_0^2t^2$, by noting that
$\int_0^\infty \!\kappa(s)ds\simeq\omega_0^2\tau_c$. Therefore $\frac1T$ should be compared to the combination of the colored-noise parameters $\omega_0^2\tau_c$.
According to Eq.~(\ref{tens_sol})
the white-noise decay is purely exponential, $e^{-j(j+1)\frac{\tau_c}T\frac t{\tau_c}}$, while in the presence of memory and for
$t\ll\tau_c$ the decay is less violent, namely $e^{-\frac12j(j+1)\frac{\tau_c}T\left(\frac t{\tau_c}\right)^2}$.  The difference is illustrated in the inset of Fig.~\ref{scales}, which shows the temporal behavior
of the damping factor at short times.
The white noise case is characterized by the omnipresent exponential decay, while the memory sets a new time-scale on which the build-up of the decoherence is slow. 
As it is pointed out in \cite{chin}, this short-time quadratic behavior is not specific to particular choice of the model, but it is an universal feature of quantum mechanics. The Markovian limit and the resulting exponential decays is an idealization which {\it always} breaks down at sufficiently short time scales where the memory-less approximation is no longer valid. 
This effect is to be colligated with a broader category of phenomena associated with quantum Zeno effect \cite{Pascazio_01,Pascazio_02,Pascazio_04,Kurizki_11,Smerzi_Zeno_12,szan_rulez}. 

Another way to control the decoherence results from an interplay between the deterministic field $\Omega$ and the fluctuating part $\boldsymbol\omega(t)$. To picture this effect, we recall
that every realization of process $\boldsymbol\omega(t)$ yields a realization of stochastic $\hat\varrho(t;\boldsymbol\omega)$ as a solution of the
equation of motion. Each realization can be viewed
in an approximate ``stroboscopic'' picture, where the field $\boldsymbol\omega(t)$ does not change much within correlation time $\tau_c$, and then jumps to another value.
The aforementioned interplay becomes clear when the evolution is viewed from the reference frame rotating around $\Omega$.
Transition to this frame does not change the $z$-component of the fluctuations which is manifested by the $\Omega$-independent decay rate $\Gamma_0$.
On the other hand, the perpendicular part rotates around the $z$-axis with frequency $\Omega$. If the period of the rotation is small in comparison to the correlation time
($\Omega\tau_c \gg 1$), it performs many revolutions between jumps.
In such a case, stochastic jumps are not noticeable anymore, different realizations become indistinguishable thus one cannot speak of fluctuations
anymore. Similar effect prevents a spinning top from toppling. This gyroscopic effect causes the decrease of $\Gamma_+$ for large $\Omega\tau_c$,
as shown in Fig.~\ref{scales}. This figure compares the damping factors $e^{-\Gamma_+}$ for three different
values of $\Omega\tau_c$.
Note that in the white-noise limit, the jumps are instantaneous which leaves no
time for the gyroscopic effect to kick in
and consequently the decay rate $\frac1T$ does not depend on $\Omega$. 


\section{Memory in noisy quantum interferometry}
\label{sec.qfi}
In the next step we examine how the build-up and gyroscopic effects can be utilized to preserve the SSN sensitivity of the interferometric parameter estimation.

\subsection{Cramer-Rao bound and the Fisher information}

The precision of the estimation of the parameter $\Omega$ from a series of $N_{\rm exp}$ experiments is bounded by the Cram\'er-Rao Lower Bound (CRLB)
\cite{braun},
\begin{equation}\label{crlb}
  \Delta^2\Omega\geqslant\frac1{N_{\rm exp}}\frac1{F_Q},
\end{equation}
where $F_Q$ is called the Quantum Fisher Information (QFI). It depends on the state of the system $\hat\varrho_S(\tau)=\overline{\hat\varrho(\tau)}$ on which the measurements are
performed and reads
\begin{equation}\label{QFI_formula}
  F_Q=2\sum_{i,j}\frac{\big|\langle i|\,\partial_\Omega\hat\varrho_S(\tau)\,|j\rangle\big|^2}{p_i+p_j}.
\end{equation}
Here, $|i\rangle$ denotes the eigenstate of $\hat\varrho_S(\tau)$ and $p_i$ is the corresponding eigenvalue.

It is important to note, that the value of $F_Q$ strongly relies upon the correlations between the particles in the system. First consider
separable states, which can be written as a convex sum of product states of $N$ particles
\begin{equation}\label{sep}
  \hat\varrho_S(\tau)=\sum_i\,p_i\,\hat\rho^{(1)}_i\otimes\ldots\otimes\hat\rho^{(N)}_i,
\end{equation}
where $0\leqslant p_i\leqslant1$ and $\sum_i p_i =1$. In this case, the QFI is bounded by the Shot-Noise Limit (SNL), $F_Q\leqslant N\tau^2$  \cite{pezze}. This bound is known to hold for unitary evolution generated by Hamiltonian of type (\ref{H_S}). Below we present a proof that this is still the case for non-unitary evolution caused by noisy Hamiltonian (\ref{ham}).

Quantum Fisher information is a convex function, hence for $\hat\varrho_S(t)$ given by a mixture (\ref{rho_mix}),  $F_Q[\hat\varrho_S(\tau)]$ is bounded by the average QFI
\begin{equation}
  F_Q[\hat\varrho_S(\tau)] \leqslant \int\!\!\mathcal D\boldsymbol\omega_0\mathcal P(\boldsymbol\omega_0) F_Q\left[\hat\varrho(\tau;\boldsymbol\omega_0)\right].
\end{equation}
Substituting the expression for the single-trajectory density matrix (\ref{single})  into Eq.~(\ref{QFI_formula}) gives
\begin{equation}\label{QFI_for_trajectory}
  F_Q\left[\hat\varrho(\tau;\boldsymbol\omega_0)\right] = 2\tau^2\sum_{i\neq j}\frac{\left|p_i-p_j\right|^2}{p_i+p_j}
  \big|\langle i|\mathbf n\cdot \hat{\mathbf J}|j\rangle\big|^2,
\end{equation}
where according to the Reference \cite{giovannetti_13}
\begin{equation}
  \mathbf n \cdot\hat{\mathbf J} = \int_0^1\!\! ds\, \hat U\!\!\left(\tau\!\cdot\! s;\boldsymbol\omega_0\right)\hat J_z\,\hat U^\dagger\!\!\left(\tau\!\cdot\! s;\boldsymbol\omega_0\right).
\end{equation} 
For separable initial state, the single-trajectory QFI from Eq.~(\ref{QFI_for_trajectory}) is bounded by $N\tau^2$ \cite{giovanetti,pezze,giovannetti_13} 
and since the probability $\mathcal P(\boldsymbol\omega_0)$ is normalized, then the average QFI from Eq.~(\ref{QFI_formula}) of an initially separable $\hat\varrho_S(\tau)$ is also bounded by $N\tau^2$. 
This concludes the proof.

When the dependence on the parameter is introduced in the system via a linear transformation as in Eq.~(\ref{ham}),
the SNL can be surpassed only when the particles are entangled, i.e. when the state cannot be written in the form (\ref{sep}).
In other words, particle entanglement of the two-mode state $\hat\varrho_S(\tau)$ is a necessary (though not sufficient) resource for having $F_Q>N\tau^2$  \cite{giovanetti,pezze}.
Therefor, quantum Fisher information can be regarded as a measure of the degree of non-classical correlations which allow for beating the SSN limit and it reaches the highest value $F_Q=N^2\tau^2$ for the maximally entangled NOON state.

On the other hand, from the point of view of the CRLB,
the QFI tells how much the state is susceptible to the imprint of the parameter $\Omega$ \cite{pezze}. The state which gives higher values of the QFI 
changes more abruptly upon the interferometric transformation,
therefore it can be used to read out even small variations of the parameter. Hence, the QFI allows to identify the optimal state for the estimation of an interferometric parameter. 
The proper choice of the probe state $\hat\varrho_S(0)$ must be accompanied by a ``good'' measurement performed on $\hat\varrho_S(\tau)$.
Such measurements can be identified formally \cite{braun} but usually are difficult to realize in the experiment. Moreover, the optimal estimation strategy, in general, will depend on the value 
of the unknown parameter $\Omega$ \cite{giovannetti_sc_natphoton}. 
This problem can be solved using some  adaptive protocols, where the information about $\Omega$ obtained from an initial, 
non-optimal estimation is used to refine the subsequent measurements.
We postpone the issue of the optimal measurements and efficiency of some more practical protocols to Section \ref{sec.cfi}.


\subsection{Fisher information in the presence of noise}

We now show, that the memory effects preserve high values of the QFI and consequently the SSN sensitivity of initially entangled states $\hat\varrho_S(0)$,
when the phase imprint is accompanied by the coupling to the environment as in Eq.~(\ref{ham}).

In the absence of noise the NOON state, used in the previous section, is maximally entangled, as $F_Q=N^2\tau^2$ reaches the ultimate Heisenberg limit
\cite{giovanetti,pezze}. As argued above, when the system in the NOON state is exposed to a noise for a long time, it reaches a completely mixed state,
which gives $F_Q=0$ and thus is utterly useless for interferometry. Nevertheless, in presence of memory effects, when $N\Gamma_+$ is small \cite{nota_small}, the QFI, which reads
\begin{equation}
  F_Q\approx\frac{N^2\big|\tau-\partial_\Omega\Gamma_-\big|^2}{1-N\Gamma_+}\,e^{-2N^2\Gamma_0-2N\Gamma_+}
  +N\Gamma_+\left(\partial_\Omega\log\Gamma_+\right)^2\label{fq}
\end{equation}
decays slower, than in the white-noise limit
\begin{equation}\label{wnQFI}
  F^{\rm(wn)}_Q=N^2\,\tau^2\,e^{-2N(N+1)\frac\tau T}.
\end{equation}
Moreover, the QFI, as seen from Eq.~(\ref{fq}), contains additional terms due to the $\Omega$-dependence of the decay rates -- the consequence of the gyroscopic effect. Because of this dependence it is expected that some knowledge about $\Omega$ can be retrieved from the observation of the evolution of the state. The second term of (\ref{fq}) quantifies how much information can be extracted in the particular example of NOON state and shows that at the short time scales it is negligible in comparison with that coming form the sole imprint.
Nevertheless, this effect is not always unimportant. For example, at longer times-scales when the off-diagonal terms of the NOON state vanish and it becomes equivalent to a mixture of the eigen-states
$\left|\frac N2,-\frac N2\right\rangle$ and $\left|\frac N2,\frac N2\right\rangle$ of the phase-imprinting operator $\hat J_z$, the QFI will remain non-zero just due to the gyroscopic effect.
We stress that this improvement is purely a memory effect,
absent in Eq.~(\ref{wnQFI}).

Notice that both in case (\ref{fq}) and (\ref{wnQFI}), the strongest source of decoherence -- which cannot be diminished by the gyroscopic effect --
comes from the $z$-component of the noise $\omega_z(t)$ which sets $\Gamma_0$ \cite{nota_N_scaling}.
\begin{figure}[hbt]
  \centering
  \includegraphics[clip,scale=0.3]{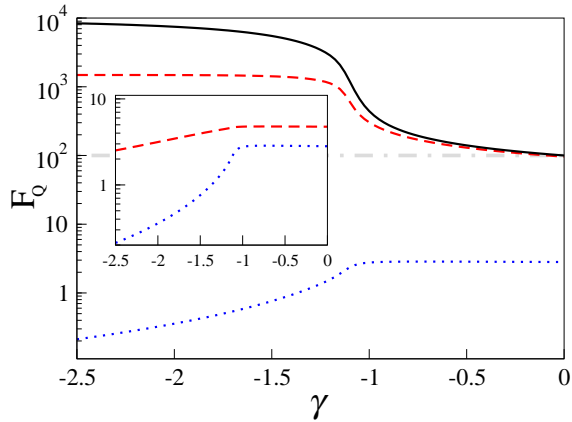}
  \caption{The QFI in units of $\tau^2$
    for different input states $\hat\varrho_S(0)$ of $N=100$ particles parametrized with $\gamma$ of the Bose-Hubbard Hamiltonian (see text for details) and with $\omega_z(t)=0$.
    The Figure compares the QFI at $\tau=0.5\tau_c$
    in absence of noise (solid black), in presence of memory (dashed red) and in the white-noise limit (dotted blue). The gray horizontal
    dot-dashed line denotes the shot-noise limit $F_Q=100$.
    The inset shows the QFI for the isotropic case in presence of memory (dashed red) and in the white-noise limit (dotted blue).
    As in Fig.~\ref{scales} the colored noise is the Ornstein-Uhlenbeck process with $\omega_0\tau_c=1$ and $\Omega\tau_c=10$.
  }    \label{fig_qfi}
\end{figure}
This is because in the Hamiltonian (\ref{ham}), $\omega_z(t)$ can be interpreted as fluctuations of the estimated parameter itself.
It is reasonable to design the experiment in such a way, that the estimated parameter is well defined.
 That is, weak and slow fluctuations along the $z$-axis is a  {\it necessary} condition for successful interferometric experiment.
In this spirit we relax the isotropic assumption (\ref{iso}) and let the strength and correlation time of $\omega_z(t)$ be different from the perpendicular components.
In particular, by setting $\omega_z=0$, we get $\Gamma_0=0$ so the strongly damping exponential factor in Eq.~(\ref{fq}) disappears.

\begin{figure}[hbt]
  \centering
  \includegraphics[clip,scale=0.6]{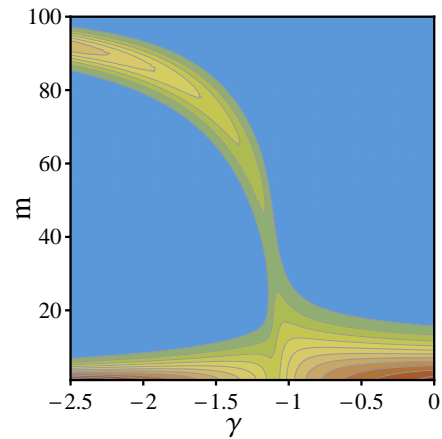}
  \caption{
    Illustration of the structure of the initial density matrices $\hat\varrho_S(0)$ generated with the Bose-Hubbard Hamiltonian. The contour plot shows the distribution
    $K_m=\sum_{j=m}^N\Big|{\rm Tr}\left(\hat\varrho(0)\hat T^{(j)}_m\right)\Big|^2$. For $\gamma\simeq0$, the density matrix consist of
    close-to-diagonal terms, so only the tensors with $m\simeq0$ contribute. When $\gamma\lesssim-1$, there is a growing contribution from large $m$'s.
  }\label{ems}
\end{figure}

To picture the impact of noise, in Fig.~\ref{fig_qfi} we plot the QFI for a wide family of usefully entangled input states $\hat\varrho_S(0)$. These are the ground states of the
two-well Bose-Hubbard (BH) Hamiltonian 
\begin{equation}\label{bh}
  \hat H_{\rm BH}=-E_J\hat J_x+U\hat J_z^2
\end{equation}
with attractive two-body interactions \cite{chwed,grond} for different values of the ratio $\gamma=U/NE_J$.
For instance, for $\gamma=0$ we have a product state called the coherent spin state (CSS),
while for $\gamma\rightarrow-\infty$ we get a NOON state. The structure of this family of states in terms of the spherical
tensors is presented in Fig.~\ref{ems}. Figure \ref{fig_qfi} shows that for $\omega_z(t)=0$, memory effects
prevent the significant loss of information and keep the particles entangled, as $F_Q>N\tau^2$. On contrary, in the white-noise case, the SSN sensitivity is lost even at short times.
Also, the inset underlines the destructive role of the fluctuations of the parameter $\Omega$.

\subsection{Symmetry and the useful entanglement}

Note that there is a striking affinity between the structure of the initial density matrices drawn in Fig.~\ref{ems} and the QFI from Fig.~\ref{fig_qfi}.
This similarity is not a coincidence and can be understood as follows. When the interferometric transformation is performed
in the absence of noise, so it is generated by the Hamiltonian (\ref{ham}) with $\boldsymbol{\omega}(t)=0$ the QFI can be bounded by
\begin{equation}
F_Q \leqslant 4
  \sum_\alpha p_\alpha\,\left(\sum_{m=0}^N m^2\sum_{j=0}^N|\langle\psi_\alpha|\hat T^{(j)}_m|\psi_\alpha\rangle|^2\right),
\end{equation}
where $\hat\varrho_{\rm S}(0) = \sum_\alpha p_\alpha |\psi_\alpha\rangle\langle\psi_\alpha|$.
 The inequality is saturated only for pure states.
The scaling of the QFI is set by magnetic number $m$, hence the presence of tensors with large $m$'s in the decomposition of the state is crucial for the SSN sensitivity.
If the state has low symmetry, the matrix elements are distributed around the diagonal -- as in the case of the CSS --
and only tensors with $m$'s close to zero are present in the state and thus the scaling is low \cite{nota_matrix_elements}.
On the other hand, if the matrix elements are stuffed in the corners --  as in the case of the NOON state --
tensors with large $m$'s will contribute
, with the largest being equal to $N$.
Consequently, one can reach up to the $N^2$ scaling of the QFI, which signals high degree of entanglement.
This reasoning shows a strict relation between the symmetry of the state, its susceptibility to the interferometric transformation and the useful entanglement.

\section{Comparison of CRLB with particular estimation strategy}
\label{sec.cfi}

Now we turn to discuss the possibility of reaching CRLB in particular estimation strategies. 

\subsection{Population imbalance}

An often applied method of estimating a parameter is based on the measurement of the population imbalance between the two output modes of the interferometer.
With ultra-cold gases, such measurement can be performed using the light imaging techniques, 
where the number of particles in spatially separated modes is obtained using the absorption imaging methods \cite{smerzi,esteve}. 
The measured difference of the number of particles $M=\frac12(N_a-N_b)$ can vary from $M=-\frac N2$ to $M=\frac N2$, while the associated probability reads
\begin{equation}\label{p_imb}
  p(M|\Omega) ={\rm Tr}\left[ \left| J,M\right\rangle\left\langle J,M\right| \hat\varrho_S(\tau)\right].
\end{equation}
Here $J=\frac N2$ and $\hat J_z\left| J,M\right\rangle=M\left| J,M\right\rangle$. 
If the measurement of the population imbalance is repeated $N_{\rm exp}\gg1$ times, the precision of the parameter estimation is bounded by 
\begin{equation}\label{classical_CRLB}
  \Delta^2\Omega \geqslant \frac 1{N_{\rm exp}}\frac 1{ F}.
\end{equation}
The quantity $F$ is called the classical Fisher information (CFI), which is given by the formula
\begin{equation}\label{cfi}
  F=\sum_{M=-\frac N2}^{\frac N2}\frac 1{p(M|\Omega)}\left(\frac{\partial p(M|\Omega) }{\partial \Omega}\right)^2.
\end{equation}
Since the QFI from Eq.~(\ref{QFI_formula}) is obtained by optimizing the precision of the parameter estimation over all possible measurements, then $F_Q\geqslant F$ holds \cite{braun}.

Note that in the absence of noise, the probability from Eq.~(\ref{p_imb}) does not depend on $\Omega$ since
\begin{eqnarray}
  p(M|\Omega)&=&{\rm Tr}\left[ \left| J,M\right\rangle\left\langle J,M\right| e^{-i\Omega\tau\hat J_z}\hat\varrho(0)e^{i\Omega\tau\hat J_z}\right]\nonumber\\
  &=&{\rm Tr}\left[ \left| J,M\right\rangle\left\langle J,M\right|\hat\varrho(0)\right].
\end{eqnarray}
In such case, the Fisher information from Eq.~(\ref{cfi}) is equal to zero and according to Eq.~(\ref{classical_CRLB}), 
the measurement of the population imbalance does not provide any information about the parameter \cite{nota_povm}. 
However, due to the gyroscopic effect the diagonal elements of the density matrix $\hat\varrho_S$ becomes $\Omega$-dependent. Hence in the presence of noise with the non-vanishing correlation time 
$F>0$. However the numerical calculations reveal that the values of $F$ remain negligible even with respect to the shot-noise level.\

\begin{figure}[hbt]
  \centering
  \includegraphics[clip, scale=0.33]{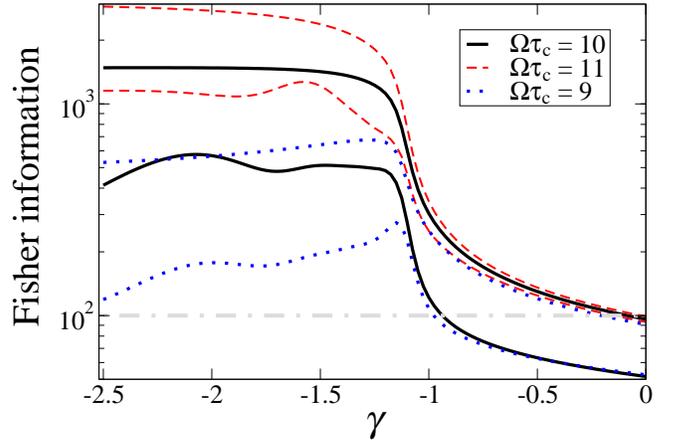}
  \caption{The Fisher information form Eq.~(\ref{cfi}) compared to the QFI for three values of $\Omega\tau_c$ plotted as a function of the input state calculated with the Bose-Hubbard Hamiltonian (\ref{bh}). 
    For each value of $\Omega\tau_c$, the upper curve (either solid, dotted or dashed)
    shows the ultimate bound of the QFI. The lower curves present the values of the Fisher information for the population imbalance estimation. 
    The plot is prepared for the same parameters as Fig.~\ref{fig_qfi}, namely $\tau=0.5\tau_c$, $N=100$ and $\omega_z(t)=0$. The horizontal dot-dashed line represents the shot-noise limit $F=100$.}
  \label{fig_bs}
\end{figure}

Nevertheless, the value of $F$ grows significantly if the information about $\Omega$ is fully exchanged between the modes before the measurement of population imbalance. 
We consider an idealized case when after the non-unitary evolution the modes are mixed by an instantaneous $\frac \pi 2$ pulse along the $x$-axis. 
Such a transformation represents a beam-splitter, a device used in light- and atom-interferometry \cite{vienna_mzi}.
In presence of this additional $\frac\pi2$-pulse, the probability from Eq.~(\ref{p_imb}) is transformed into
\begin{equation}
  p(M|\Omega) ={\rm Tr}\left[ \left| J,M\right\rangle\left\langle J,M\right| e^{-i\frac\pi2\hat J_x}\hat\varrho_S(\tau)e^{i\frac\pi2\hat J_x}\right].
\end{equation}
In the absence of noise this probability gives $F=F_Q$ if the initial state satisfies ${\rm Tr}\left[\hat J_z\hat\varrho_S(0)\right]=0$, see \cite{opt_measurements}.
Such states are called path-symmetric, and among them are the ground states of the BH Hamiltonian (\ref{bh}) for all $\gamma$. 
When the noise is present, the measurement of the population imbalance is not optimal anymore, i.e. $F<F_Q$ as shown in Fig.~\ref{fig_bs}. The Fisher information is much smaller then the QFI 
despite strong suppression of decoherence due to memory effects. Still, as Fig.~\ref{fig_bs} shows, the discussed protocol can give sub shot-noise sensitivity with entangled input states.
 

\subsection{One-body density}

Finally, we consider a strategy of the parameter estimation from the one-body density \cite{chwed}, which is well suited for interferometers based on ultra-cold atom systems. 
In the scenario we examine the beam-splitter, which is often difficult to implement in the atomic systems, is replaced by the free expansion of the two clouds initially localized around the 
minima of the double-well potential. 
After the sufficiently long expansion time, the clouds overlap and form the interference pattern in the far-field regime. 
The positions of individual atoms can be precisely detected using modern techniques such as the micro-channel plate \cite{plate}, 
tapered fiber \cite{fiber}, the light-sheet method \cite{light_sheet} or the atomic fluorescence form the lattice \cite{lattice}.
The acquired data, in principle, gives access to atom-atom correlations of all orders.
The position of the observed interference fringes depends on $\Omega$, and provides the information about the interferometer.
The estimation strategy based on measurements of $N$th-order correlation function is optimal for a wide spectrum of input states $\hat\varrho_S(0)$, 
because it utilizes the information about the whole density matrix of the system \cite{chwed_N_corr_10,chwed_N_corr_11}. 
However, even with small number of particles, the measurement of the $N$th-order correlation function would require substantial experimental effort. 
For this reason we limit our discussion to estimation from first correlation function - the one-body density.

The general scheme of the estimation from the parameter-dependent density $\rho$ is following. In every single shot of the experiment the positions of atoms forming the interference pattern are recorded. 
The experiment is repeated $N_{\rm exp}$ times and the density $\rho(x|\tilde\Omega)$ is fitted to the averaged data points. Here, $\tilde\Omega$ is a free parameter, which is determined from the least-squares
formula. If this procedure is performed large number of times, the averaged $\tilde\Omega$ tends to the true value of the parameter $\Omega$, meaning that the estimator $\tilde\Omega$ is unbiased. 

\begin{figure}[hbt]
  \centering
  \includegraphics[clip, scale=0.33]{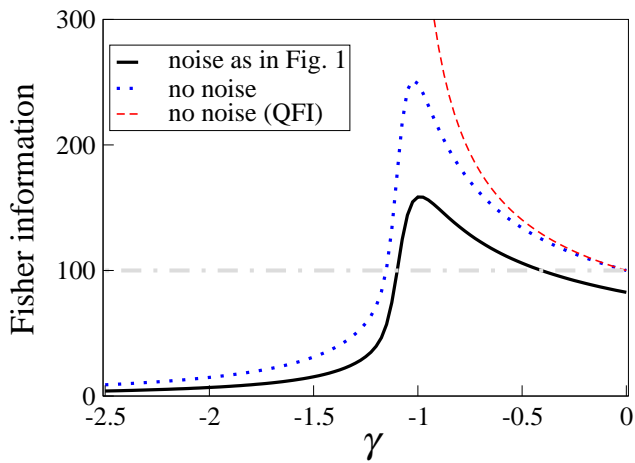}
  \caption{The Fisher information from Eq.~(\ref{CFI_1body}) for different input states calculated with the Bose-Hubbard Hamiltonian (\ref{bh}). The solid black line is the $\tilde F$ for the
    in presence 
    of noise. We use the same model of the Ornstein-Uhlenbeck process as in Fig.\ref{scales} with $\Omega\tau_c=10$ for $\tau=0.5\tau_c$. 
    The dotted line is the maximal value of $\tilde F$ in the absence of noise. For reference, the ultimate bound of the QFI is shown with the red
    dashed line. The dot-dashed gray line represents the shot noise limit $\tilde F=100$.}
  \label{fig_den}
\end{figure}

In this protocol, the value of the estimator is derived from the one-body density. In consequence, the sensitivity, which is the variance of the estimator, 
might depend on the variance of the density -- the second order correlation function. A rigorous derivation of the sensitivity \cite{chwed} gives
\begin{equation}
  \Delta^2\tilde\Omega=\frac1{N_{\rm exp}}\frac1{F_1+F_2}\equiv\frac1{N_{\rm exp}}\frac1{\tilde F}.
\end{equation}
In the above expression, $F_1$ is the one-body Fisher information and reads
\begin{equation}\label{f1}
  F_1=\int dx\frac1{\rho(x|\Omega)}\left(\frac{\partial\rho(x|\Omega)}{\partial\Omega}\right)^2.
\end{equation}
The other component of the sensitivity is related not only to the density, but also to the second-order correlation function 
\begin{equation}\label{f2}
  F_2=\frac1{F_1}\int\!\! dx\!\!\int\!\! dy\,\frac{\partial\rho(x|\Omega)}{\partial\Omega}\frac{\partial\rho(y|\Omega)}{\partial\Omega}\,g^{(2)}(x,y|\Omega).
\end{equation}
In order to evaluate the integrals (\ref{f1}) and (\ref{f2}), we introduce the two-mode field operator $\hat\Psi(x)=\psi_a(x)\hat a+\psi_b(x)\hat b$. In the far-field regime, the mode functions 
$\psi_a(x)$ and $\psi_b(x)$ fully overlap and only differ by the phase. The density is defined as
\begin{equation}
  \rho(x|\Omega)={\rm Tr}\left[\hat\Psi^\dagger(x)\hat\Psi(x)\hat\varrho_S(\tau)\right],
\end{equation}
while the second-order correlation function reads
\begin{equation}
  g^{(2)}(x,y|\Omega)=\frac{{\rm Tr}\left[\hat\Psi^\dagger(x)\hat\Psi^\dagger(y)\hat\Psi(y)\hat\Psi(x)\hat\varrho_S(\tau)\right]}{\rho(x|\Omega)\rho(y|\Omega)}.
\end{equation}
These two expectation values can be calculated analytically and inserted into Equations (\ref{f1}) and (\ref{f2}) give a rather complicated expression
\begin{equation}
  \tilde F=\frac{\langle\hat J_x\rangle^2(\alpha+\mu\beta)^2}
         {\frac N4(\alpha+\beta)\mu+\langle(\alpha\hat J_y-\beta\hat J_x)^2\rangle-\alpha^2\langle\hat J_y\rangle^2-\beta^2\langle\hat J_x\rangle^2}\label{CFI_1body}.
\end{equation}
Here $\alpha=\partial_\Omega\left(\Omega\tau-\Gamma_-\right)$ and $\beta= \frac1{\mu}\partial_\Omega\Gamma_+$ are the rates of changes of imprinted phase and decay rates.
Also, $\mu=\sqrt{1-\frac{2}{N}\langle\hat J_x\rangle}$ is related to the visibility of the interference fringes. In the above expression, all the average values are calculated with the initial state, 
so for instance $\langle\hat J_x\rangle={\rm Tr}\left(\hat J_x\hat\varrho_S(0)\right)$.

Figure \ref{fig_den} shows $\tilde F$ for the family of ground states of the BH Hamiltonian for various values of $\gamma$. Although the noise suppresses the value of $\tilde F$, it can still give sub-shot
noise sensitivity. The characteristic drop of $\tilde F$ around $\gamma=-1$ is related to the decrease of the visibility of the fringes in the one-body density due to transition from the coherent-like
to the NOON-like states of the BH Hamiltonian.

Equation (\ref{CFI_1body}) depends on the two lowest moments of the angular momentum operators which probe the parts of density matrix spanned by the $m=0,1,2$ tensors.  
Therefore, the value of $\tilde F$ is high for those states, which have low symmetry. According to Fig (\ref{ems}), those are the states for $-1\lesssim\gamma \leq 0$, which
have their matrix elements focused mostly around the diagonal. On the other hand, $\tilde F$ drops significantly for highly symmetric states. 
In order to probe the parts of density matrix spanned by tensors of larger $m$-s, the estimation strategy should be extended to measurements of higher correlation functions. 
According to results of \cite{chwed}, CFI for estimation from $n$th correlation function depends on up to $2n$th moments of angular momentum operators, 
which in turn are spanned by spherical tensors of $m$-s reaching $2n$.

\section{Conclusions and acknowledgments}
\label{sec.conc}
We have demonstrated how the memory in an $N$-qubit system, which is a subject to external noise, can be employed to suppress the entanglement decay.
We have argued that this effect has potential application to quantum interferometry. In presence of memory, the deterministic evolution -- which imprints the parameter-$\Omega$
dependence on the state
-- couples to the non-unitary evolution generated by the fluctuations. As a result, the SSN sensitivity is retained for longer times -- not only due to the decoherence slow-down
but also thanks to additional information about $\Omega$ present in the decay rates. This observation is valid not only for the ultimate bound of the precision but also applies to some particular
estimation protocols.

P. Sz. acknowledges the Foundation for Polish Science International Ph.D. Projects Program co-financed by the EU European Regional Development Fund.
This work was supported by the National Science Center grant no. DEC-2011/03/D/ST2/00200.
M. T. was supported by the National Science Center grant.



\begin{thebibliography}{50}
\bibitem{nota_cut}{
  The numerical simulations reveled that the cumulants of higher orders must be taken into account in quasi-static regime when $\tau_c\to\infty$, 
  but only for times following the period of evolution when the average reaches it first minimum, which is very close to zero. 
  At this point the higher order cumulants take over and ``revive'' the average to $\sim\frac 1{2j+1}\big\langle \hat T^{(j)}_m(0)\big\rangle$, 
  while second order approximation would result in progressing decay to zero. The time at which the average reaches the minimum is roughly estimated as 
  $t\sim\left(\int_0^\infty ds \kappa(s)\cos[\Omega s]\right)^{-1}$}
\bibitem{nota_small}{
  Even if the order of magnitude of $\Omega$ is not know and it is impossible to assess whether $N\Gamma_+\ll 1$, this criterion suggest a strategy where the additional, 
  well controlled field $\Omega_0$ is applied. Then one is tasked with estimation of a total field $\Omega_{\rm{tot}}=\Omega+\Omega_0$ instead of $\Omega$, 
  hence the useful entanglement can still be utilized to go beyond the shot-noise while maintaining strong gyroscopic effect.}
\bibitem{nota_N_scaling}{
  According to Eq. (\ref{fq}) the decay caused by the fluctuations in the $z$-direction are amplified by $N^2$ as opposed to $N$, as reported in \cite{chin}. 
  This important discrepancy between the two results is due to the fundamental difference in the adopted physical models. 
  In our case qubits are identical bosons, hence they can only experience the common noise. 
  In \cite{chin} qubits are distinguishable and can couple to local, uncorrelated sources of noise. 
  Consequently the net decay rate is simply a sum of decay rates of each qubit -- hence the amplification with the number of particles.}
\bibitem{nota_matrix_elements}{
  According to Wigner-Eckhart theorem the matrix element of the normalized spherical tensor is given by $\langle J,M|\hat T^{(j)}_m|J,M'\rangle = \langle J,M' ; j,m| J, M \rangle\sqrt{\frac{2j+1}{2J+1}}$. 
  The Clebsch-Gordan coefficient does not vanish if and only if the condition $M'+m = M$ is satisfied. Hence, the matrix of $\hat T^{(j)}_m$ 
  is sparse and the non-zero elements are located in a band positioned $m$ units away from the diagonal.}
\bibitem{nota_povm}{
  In more technical terms, positive operator valued measure (POVM) with elements $\hat\Pi_{M} = \left|J,M\right\rangle\left\langle J,M\right|$ probe only the diagonal of the density matrix, 
  which is spanned by $m=0$ spherical tensors. These tensors are not affected by unitary rotations such as (\ref{H_S}) (see Eq. (\ref{tens_sol}), where the phase $\Omega\tau$ is multiplied by $m$), 
  thus in absence of noise $F=0$.}
\end{thebibliography}
\end{document}